\documentclass{aa} 
\usepackage{txfonts}  
\usepackage{graphicx}  
\usepackage{natbib}  
\usepackage{here}  
\usepackage[normalem]{ulem}
\bibpunct{(}{)}{;}{a}{}{,} 
\newcommand{\nc}{\newcommand}  
\newcommand{\micron}{\mbox{$\mu$m}}  
\newcommand{\msun}{M$_{\rm \odot}$}  
\newcommand{\lsun}{L$_{\rm \odot}$}  
\newcommand{\ea}{et~al.}  
  
\newcommand{\CI}{[\ion{C}{i}]}  
\newcommand{\CII}{[\ion{C}{ii}]}  
\newcommand{\NII}{[\ion{N}{ii}]}  
\newcommand{\HII}{\ion{H}{ii}}  
\newcommand\arcdeg{\mbox{$^\circ$}}%
\nc{\twCO}{$^{12}$CO}  
\nc{\thCO}{$^{13}$CO}  
\nc{\CeiO}{C$^{18}$O}  
\nc{\cmcub}{\mbox{cm$^{-3}$}}  
\nc{\cmsq}{\mbox{cm$^{-2}$}}  
\nc{\Kkms}{\mbox{K~km~s$^{-1}$}}  
\nc{\kms}{\mbox{km~s$^{-1}$}}  
\nc{\ext}{\mbox{T$_{\rm ex}$}}  
\nc{\nhtwo}{\mbox{N(H$_2$)}}
\nc{\Cone}{\mbox{[\ion{C}{i}]$^3$P$_1$--$^3$P$_0$}}  
\nc{\Ctwo}{\mbox{[\ion{C}{i}]$^3$P$_2$--$^3$P$_1$}}  
\nc{\cbyco}{\mbox{C/CO}}

\begin{document}  
  
\title{Study of Photon Dominated Regions in Cepheus~B}

\author{B. Mookerjea\thanks{\emph{Present address}: Department of  
Astronomy, University of Maryland, College Park, MD 20742, USA}  
 \and C. Kramer  
  \and M. R\"ollig
  \and M. Masur}
  
\institute{KOSMA, I. Physikalisches Institut, Universit\"at zu  
K\"oln, Z\"ulpicher Strasse 77, 50937 K\"oln, Germany }
  
\offprints{B. Mookerjea, \email bhaswati@astro.umd.edu}  
  
\date{Received /Accepted }  
  
\titlerunning{PDRs in Cepheus~B }
\authorrunning{Mookerjea et al.}  
  
\abstract  
{}{ The aim of the paper is to understand the emission from the photon
dominated regions in Cepheus~B, estimate the column densities of
neutral carbon in bulk of the gas in Cepheus~B and to derive
constraints on the factors which determine the abundance of neutral
carbon relative to CO.} {This paper presents
15\arcmin$\times$15\arcmin\ fully sampled maps of \CI\ at 492~GHz and
\twCO\ 4--3 observed with KOSMA at 1\arcmin\ resolution.  The new
observations have been combined with the FCRAO \twCO\ 1--0, IRAM-30m
\thCO\ 2--1 and \CeiO\ 1--0 data, and far-infrared continuum data from
HIRES/IRAS. The KOSMA-$\tau$ spherical PDR model has been used to
understand the \CI\ and CO emission from the PDRs in Cepheus~B and to
explain the observed variation of the relative abundances of both
C$^0$ and CO.} {The emission from the PDR associated with Cepheus~B is
primarily at $V_{\rm LSR}$ between -14 and -11~\kms. We estimate about
23\% of the observed \CII\ emission from the molecular hotspot is due
to the ionized gas in the \HII\ region. Over bulk of the material the
C$^0$ column density does not change significantly,
$(2.0\pm1.4)\,10^{17}$\,cm$^{-2}$, although the CO column density
changes by an order of magnitude.  The observed \cbyco\ abundance
ratio varies between 0.06 and 4  in Cepheus B. We find an
anti-correlation of the observed \cbyco\ abundance ratio with the
observed hydrogen column density, which holds even when all
previous observations providing \cbyco\ ratios are included.  
Here we show that this observed variation of C/CO abundance with total
column density can be explained only by clumpy PDRs consisting of an
ensemble of clumps. At high H$_2$ column densities high mass clumps,
which exhibit low C/CO abundance, dominate, while at low column
densities, low mass clumps with high C/CO abundance dominate.}{}  
  
\maketitle  
  
\section{Introduction}  
  
Photon Dominated Regions (PDRs) are predominantly neutral, atomic and
molecular regions where the physical and chemical processes are
dominated by Far Ultraviolet (FUV) radiation ({\em cf.}
\citet{sternberg1995,hollenbach1997}) and the major carbon-bearing
species changes from C$^+$ through C$^0$ to CO. While FUV photons (6
eV$< h\nu <13.6$~eV) are primarily responsible for the heating of the
cloud surfaces via photoelectric effect on dust grains, at larger depths
of about 10~mag into the clouds cosmic ray induced heating dominates.
Owing  to the clumpy nature of the molecular clouds, contrary to the
initial  ideas, PDRs are not strictly confined to the surface of the
molecular  clouds. Rather the extended \CII\ and \CI\ emission observed
from  allover the molecular clouds suggest formation of PDRs deep inside
the  molecular clouds, at the surfaces of clumps inundated with
FUV  photons escaping into the cloud. Further, the chemical abundances
of several key species like e.g., CN and HCN depend to a large extent on
the abundances of C$^0$ and C$^+$ in the gas phase \citep{boger2005}.
PDRs may therefore be seen as the most  ubiquitous phase of the ISM,
observations of which together with proper  modeling provide insight
into the physical conditions within the  molecular clouds.  On a more
global scale, PDRs provide direct insight  into the chemical evolution
of the ISM deluged with FUV photons and PDR models suggest
radiation-induced feedback mechanisms  that may regulate star formation
rates and the column density of gas  through giant molecular clouds.
  
To the south of the Cepheus OB3 association of early-type stars lies the
giant Cepheus molecular cloud, at a distance of about 730~pc
\citep{blaauw1964}. Cepheus~B, with a size of 4~pc, is the hottest
\twCO\ component \citep{sargent1977,sargent1979} and is located near the
northwestern edge of the cloud, adjacent to the \HII\ region S155
\citep{sharpless1959}. The interface between the molecular cloud and the
OB stars, is delineated by the optically visible \HII\ region  S155,
whose very sharp edges clearly indicate the presence of ionization
fronts bounding the dust/molecular cloud. The OB association itself
consists of two subgroups of different ages, with the youngest lying
closer to the molecular cloud \citep{sargent1979}.  \citet{felli1978}
presented radio continuum observations which confirmed that the
molecular cloud Cepheus~B is surrounded by an ionization front driven by
the radiation field of the bright stars that are part of the youngest
generation of the Cepheus~B OB3 association.  The structure of the
diffuse ionized gas and its energy balance can be accounted for by the
UV radiation coming from the brightest members of the OB3 association
(viz., HD217086 and 217061). The physical association between the
Cepheus~B molecular cloud and the S155 \HII\ region was also confirmed
by  the H$_2$CO and recombination line observations of
\citet{panagia1981}.  \citet{minchin1992} used a combination of low-J CO
and far-infrared  (FIR) continuum observations to show that the
Cepheus~B molecular cloud  terminates at its north-west edge with a
hotspot. The hotspot lies  1\arcmin\ to the east of the peak in CO
column density. The high temperature of  the hotspot and the associated
FIR emission cannot be explained by only  considering the stars in the
OB3 association. It was further proposed by  \citet{minchin1992} and
later confirmed using radio and near-infrared  (NIR) continuum
observations by \citet{testi1995} that the hotspot is  indeed energized
by a highly reddened main sequence star (NIR-A) of  spectral type
B1-B0.5, together with a small cluster of young stars. 

In a first study with the KOSMA 3m telescope, \citet{beuther2000}
mapped the Cepheus B cloud in 2--1 and 3--2 transitions of $^{12}$CO,
\thCO, and \CeiO\  at 2\arcmin\ resolution and used PDR models of
\citet{stoerzer2000} to interpret the emission at four selected
positions. The hotspot emission indicated the presence of shocks. 
Based on the derived local volume densities of $\sim 2~10^4$~\cmcub,
and average volume density less than 10$^3$~\cmcub,
\citet{beuther2000} concluded that Cepheus~B is highly clumped with
clumps filling  only 2\% to 4\%  by volume of the cloud. In this
work \CI\ at 492~GHz was detected from two selected positions and
detection of \CI\ emission from deep inside the cloud was also put
forward as an additional evidence for the clumpiness of Cepheus~B.

In clumpy or homogeneous PDR scenario both \CI\ and \twCO\
4--3 emission arise from regions closer to the surfaces of the FUV
irradiated clumps or molecular cloud surfaces respectively.  Thus the
\CI\ and \twCO\ 4--3 emission trace similar regions.  While \CI\
emission necessarily implies photodissociation of molecular CO, the
higher kinetic temperature (the upper energy level being $56$~K) required
to excite the \twCO\ 4--3 line suggest origin of this emission from
the outer surfaces of dense (since the critical density of the \twCO\
4--3 transition for collisions with H$_2$ is $\sim 10^5$~\cmcub)
clumps embedded in a UV radiation field.  Thus in order to trace the
PDR we have observed largescale ($15'\times15'$) fully-sampled maps of
the \CI\ and \twCO\ 4--3  emission from the Cepheus~B molecular cloud.
At a common resolution of 1\arcmin, this dataset has been combined
with FCRAO \twCO\ 1--0, IRAM-30m \thCO\ 2--1 and \CeiO\ 1--0 data, and
FIR continuum data from HIRES/IRAS.  This allows for a
detailed analysis of the PDRs in Cepheus B, since  observed
transitions span a substantial range of critical densities and
temperatures (Table~\ref{tab_exc}).

\begin{table}[hbt]
\caption{Parameters of the observed transitions used to study the PDRs
in Cepheus~B.
\label{tab_exc}}
\begin{center}
\begin{tabular}{llll}
\hline
\hline
Species & Transition & $E_{\rm u}$/k & $n_{\rm cr}$ \\
&& (K) &  (\cmcub)\\
\hline
\CI\ & $^3P_1\rightarrow^3P_0$ & 23.6 & 1~10$^3$\\
\twCO\  & 1--0 & 5.6 & 4.8~10$^3$ \\
\thCO\  & 2--1 & 16.8 & 3.2~10$^4$ \\
\twCO\ &  4--3 & 56  & 2.6~10$^5$ \\
\hline
\end{tabular}
\end{center}
\end{table}

\section{Datasets}  
\begin{figure*}[htb]
\begin{center}  
\includegraphics[angle=0,width=16.0cm,angle=0]{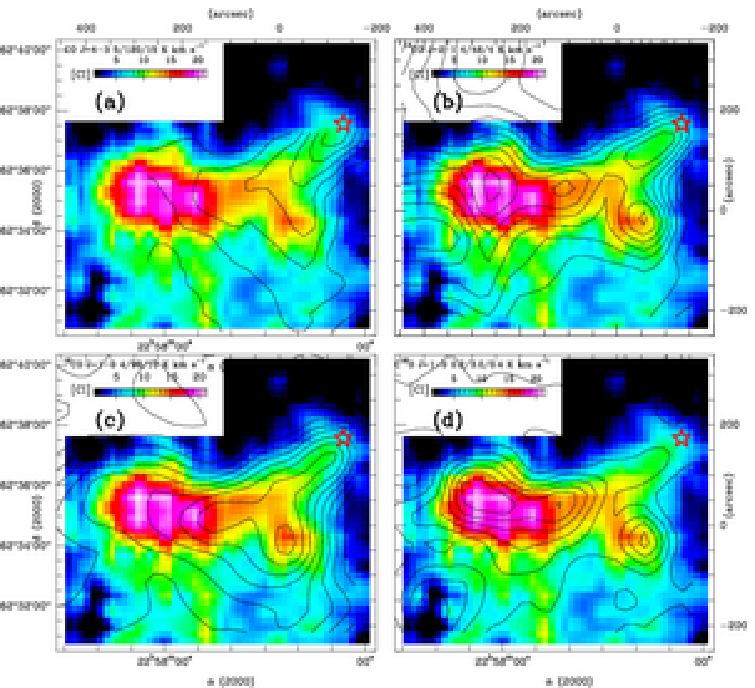}  
\caption  
{Velocity-integrated intensity maps of \CI\ $^3$P$_1$-$^3$P$_0$  emission
({\em color}) overlayed with contours of {\bf \em (a)} \twCO\ 4--3, {\bf
\em (b)} \thCO\ 2--1, {\bf \em (c)} \twCO\ 1--0 and {\bf \em (d)} \CeiO\
1--0 intensities.  All maps are integrated over the range V$_{\rm LSR}$=-20
to -5 km~s$^{-1}$.  Colorscale of the \CI\ map and contour levels of all
other line intensities are indicated at the top left corner of each panel.
The red star denotes the position of the {\em hot spot}. All data are
shown at a common resolution of 1\arcmin.
\label{cepb_intmap}}  
\end{center}  
\end{figure*}   
  
\subsection{KOSMA Observations of PDR tracers: \CI\ and \twCO\ 4--3}  
  
We have used the KOSMA 3m submillimeter telescope on Gornergrat,  
Switzerland \citep{winnewisser1986,kramer1998} to observe the emission  
of fine-structure line of neutral carbon at 492~GHz (609~\micron,  
$^3$P$_1$--$^3$P$_0$; hereafter \CI) and J = 4--3 transition of \twCO\  
emission at 461 GHz from the PDRs associated with Cepheus~B. 
We used the Submillimeter Array Receiver for Two frequencies (SMART)  
\citep{graf2002} on KOSMA for these observations. SMART is a  
dual-frequency eight-pixel SIS-heterodyne receiver capable of observing  
simultaneously in the two atmospheric windows centered around  
650~\micron\ and 350~\micron. The IF signals were analyzed with  
array-acousto-optical spectrometers with a spectral resolution of  
1.5~MHz \citep{horn1999}.   
  
The observations were performed in the position-switched On-The-Fly mode
between November 2004 and March 2005.  The HPBW of the \CI\ and \twCO\
4--3 observations are 60\arcsec\ with a beam efficiency of $\eta_{\rm
mb}= 50$\%. Atmospheric calibration was done by measuring the
atmospheric emission at the OFF-position to derive the opacity
\citep{hiyama1998}. Sideband imbalances were corrected using standard
atmospheric models \citep{cernicharo1985}. All data presented in this
paper are in units of main beam temperature (T$_{\rm mb}$), calculated
from the observed calibrated antenna temperature (T$_A^\ast$) using the
derived beam efficiencies ($\eta_{\rm mb}$) for the relevant  telescope,
T$_{\rm mb}$ = T$_{A}^\ast/\eta_{\rm mb}$. From observations of reference
sources we estimate the relative calibration accuracy to be $\sim 15$\%.
Data was reduced using the GILDAS\footnote{\tt
http://www.iram.fr/IRAMFR/GILDAS} astronomical data reduction package.   
  
\subsection{Further CO data}   
  
We have complemented the \CI\ and \twCO\ 4--3 datasets for Cepheus~B
with low-J CO observations available in literature. We have obtained the
\twCO\ 1--0 dataset from the FCRAO outer Galaxy survey \citep{heyer1998}
and \thCO\ 2--1 and \CeiO\ 1--0 datasets observed with the IRAM~30m by
\citet{ungerechts2000} in 1995.  The beam efficiency, $\eta_{\rm mb}$
(defined as B$_{\rm eff}$/F$_{\rm eff}$) used to derive the T$_{\rm mb}$
scale are 0.45, 0.48 and 0.75 for the \twCO\ 1--0, \thCO\ 2--1 and \CeiO\ 1--0
lines respectively \citep{heyer1998,iramn1994}. All datasets have been
smoothed to a resolution of 1\arcmin\  for comparison with our
observations.  These CO datasets provide deeper  insight into the
physical conditions of the regions emitting \CI\ and  \twCO\ 4--3
emission. \twCO\ 1--0 being optically thick, is most likely  to be
arising from closer to the PDR surface. The \thCO\ 2--1 and \CeiO\  1--0
on the other hand are most likely to be optically thin and hence  arise
from deeper inside the clumps, and hence are crucial for  ascertaining
the physical parameters (like the optical extinction)  characterizing
the emitting regions.  
  
\subsection{HIRES processed IRAS maps}  
  
Dust continuum emission  accounts for about 98\% of the cooling of the
molecular clouds.  Here we have used FIR continuum maps to
estimate the intensity of the FUV radiation field  crucial for the
chemical and physical structure of the PDRs.  We have obtained
HIRES-processed \citep{aumann1990} IRAS maps of Cepheus~B at 60 and
100~\micron\ from the Infrared Processing and Analysis  Center,
Caltech.  These dust continuum maps have angular resolutions 
(1\farcm5) comparable to our sub-millimeter datasets.

\section{Observational Results}  
  
\subsection{Integrated Intensity Maps}

Figure~\ref{cepb_intmap}(a) shows the integrated intensity maps of \CI\
and \twCO\ 4--3  emission from Cepheus~B. The maps are centered at  
$\alpha=$ 22$^h$57$^m$25\fs7  $\delta = $ 62\arcdeg34\arcmin40\farcs7
(J2000).  The \twCO\ 4--3 emission peaks at the position of the hotspot,
and drops off significantly towards the south-east, and does not show
much substructures. The \CI\ emission in contrast though rather weak at
the position of the hotspot,  has almost uniform intensity over the
ridge extended along the east-west  direction with an isolated peak to
the south-west (below the hotspot)  and an extended high intensity peak
deep into the molecular cloud to the  east of the map.  The somewhat
different distribution of \twCO\ 4--3 and  \CI\ emission suggest that
while the \twCO\ 4--3 emission  traces higher temperature and densities
as in the hotspot, the \CI\ emission being optically thin traces the
embedded PDR surfaces of the  molecular material with higher column
density better.  The features seen  in the \twCO\ 4--3 map agree
reasonably well with the features seen in  the \twCO\ 3--2 map by
\citet{beuther2000}.  
  
Figures~\ref{cepb_intmap} (b), (c) and (d) show overlays of the observed
\CI\ emission with the \thCO\ 2--1, \twCO\ 1--0 and \CeiO\ 1--0
intensities respectively. The \CI\ and  \thCO\ 2--1 appear to be almost
concomitant tracing similar features in the Cepheus~B cloud.  To the
west of the map the \twCO\ 1--0 emission is similar in parts with the
\CI\ emission. However the \twCO\ 1--0 intensities drop off more
drastically towards the east, which is similar to the features seen in
the \twCO\ 4--3 map. This is probably due to the higher column densities
to the east, as manifested by the peaks in the \thCO\ and \CeiO\ maps.
The lower \twCO\ 4--3 emission to  the east, stem from a combination of
lower densities and kinetic  temperatures. The \CeiO\ 1--0 emission,
tracing higher column densities, looks  markedly different from the
\twCO\ 4--3 emission and almost similar to  the \CI\ emission except for
the region close to the hotspot where, the  \CeiO\ emission is almost
absent. The faint \thCO\ 2--1 emission and  almost no \CeiO\ 1--0
emission from the hotspot is consistent with the  results of
\citet{minchin1992}, who found a distinct lack of \thCO\  emission at
the position of the hotspot.  
  
\subsection{Velocity Structure}  
  
Previous low-J CO observations showed that the Cepheus~B emission  
consists of two major velocity components: one, the north-western part  
of the cloud closer to the hot spot emits at $\sim -15$~\kms,  
confirming the association of the hot spot with S155 having the same  
system velocity and second, the south-eastern part emitting at $\sim  
-12$~\kms\ \citep{beuther2000,minchin1992}.   
  
\begin{figure}[h]  
\begin{center}  
\includegraphics[angle=0,width=9.0cm,angle=0]{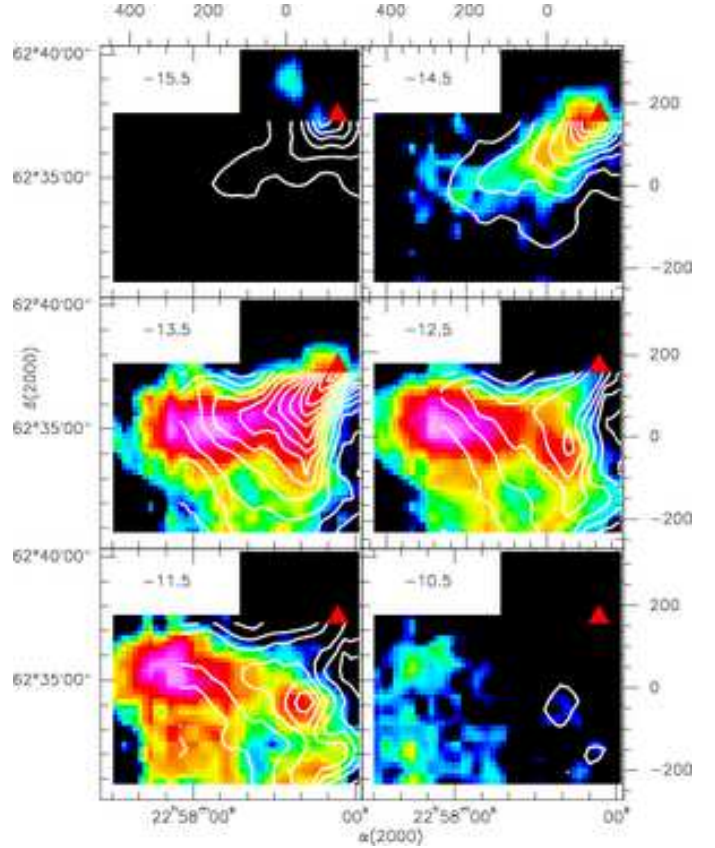}  
\caption{Overlay of \CI\ (color) and \twCO\ 4--3 (contours) velocity   
channel maps of Cepheus~B.  The colorscale ranges between 1 and  
11.6~\Kkms\ and the \twCO\ 4--3 contours are between 5 and 120 \Kkms\ in  
steps of 15~\Kkms.  
\label{cepb_chanmap}}  
\end{center}  
\end{figure}

An overlay of the  \CI\ and \twCO\ 4--3 channel maps
(Fig.~\ref{cepb_chanmap}) shows that bulk of the cloud emits between -14
and -11 \kms, although substructures contributing to the \twCO\ 4--3
emission change from the hotspot to the clump to the south-west of the
map. The emission from the hotspot is slightly blue-shifted, between -16
and -14~\kms.  At -15.5~\kms, we detect in \CI\ a clump to the  extreme
north of the map. At -10.5~\kms\ the \twCO\ 4--3 emission from  the
cloud is too low and the \CI\ emission is restricted mainly to the  east
of the ridge. It is interesting to note that at -11.5~\kms\ the \CI\ and
\twCO\ 4--3 emission from the south-western clump are exactly
concomitant. 
  
\begin{figure}[hbt]  
\begin{center}  
\includegraphics[angle=0,width=7.0cm,angle=0]{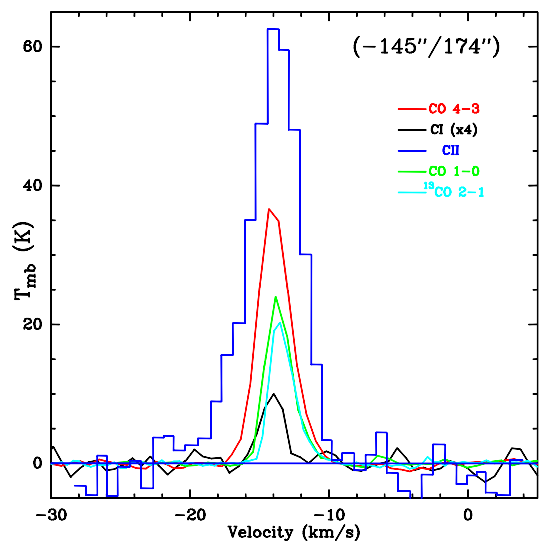}  
\caption {Observed CI, \twCO 4--3, \twCO 1--0, \thCO 2--1  \& \CII\ spectra 
at  the position of the molecular hotspot in
Cepheus~B. The \CII\ spectrum is adopted from \citet{boreiko1990}. The ratio 
of integrated intensities \twCO 4--3/\twCO 1--0 
at this position is  1.9. All data are shown at a common resolution of
1\arcmin, except for the \CII\ data which has a resolution of 43\arcsec.
\label{fig_posspec}}  
\end{center}  
\end{figure}   

\begin{table}[hbt]
\caption{Velocity components derived from gaussian fitting of spectra at 
the position of the hotspot in Cepheus~B.
\label{tab_hotspot}}
\begin{center}
\begin{tabular}{lll}
\hline
\hline
Line & V$_{\rm LSR}$ & FWHM \\
& (\kms) & (\kms) \\
\hline
\CI\ & -14.1 & 1.9\\
\twCO\ 4--3 & -13.9 & 2.8 \\
\twCO\ 1--0 & -13.6 & 2.2 \\
\thCO\ 2--1 & -13.4 & 2.1\\
\CII\ Comp 1& -13.6 & 3.8 \\
\CII\ Comp 2& -16.5 & 6.0 \\
\hline
\end{tabular}
\end{center}
\end{table}

Figure~\ref{fig_posspec} shows the \CI, CO  and \CII\ 158~\micron\ spectra
observed at the position of the hotspot. The \CII\ spectrum was observed
by \citet{boreiko1990} using the Kuiper Airborne Observatory with a
resolution of 43\arcsec, similar to the resolution of the other spectra
(60\arcsec).  The velocity resolution is 0.8~\kms. It is one of the very
few heterodyne spectra of the \CII\ line taken so far.  These spectra
show that at the position of the hotspot the PDR emission is dominated
by the velocity component at V$_{\rm LSR}$ of -13.8~\kms\ and the
lineshapes with the exception of \CII\ are all primarily gaussian.  The
\CII\ line at the hotspot position appears to be the broadest, with
traces of an extended blue wing. \citet{boreiko1990} have explained the
\CII\ emission from the hotspot position to be from two velocity
components, the main narrow component is centered at -13.6~\kms\ with a
width of 3.8~\kms, while the blue component is centered at -16.8~\kms\
and has a width of 6.0~\kms.  Table~\ref{tab_hotspot} presents results
of fitting Gaussian profiles to the different spectral lines observed at
the hotspot. Broadly we find that both \CI\ and CO emission arise around
$V_{\rm LSR}$ of -13.8~\kms\ and have linewidths $\sim 2.2$~\kms. We
have reproduced the gaussian fits to the \CII\ spectra obtained by
\citet{boreiko1990} using the same parameters.  The broader blue
component at -16.5~\kms\ does not appear to be associated with the
emission from the PDR/molecular cloud at all, rather the velocity
matches the V$_{\rm LSR}$ shown by H$\alpha$ emission from the H {\sc
ii} region S155 \citep{boreiko1990}. Based on the gaussian fit to the
\CII\ spectrum, we estimate $\sim 77$\% of the measured \CII\ emission
arises from the PDR and the remaining from the ionized gas.  The
estimated relative contribution of PDRs to \CII\ emission from nearby
galaxies using the \NII\ 122~\micron\ is $\sim$ 70--85\%
\citep{malhotra2001,kramer2005}, roughly consistent with the above
result for Cepheus~B.

\section{LTE Analysis}  
  
\subsection{The CO(4-3)/CO(1-0) distribution}  
  
\begin{figure*}[hbt]  
\begin{center}  
\includegraphics[angle=0,width=8.0cm,angle=0]{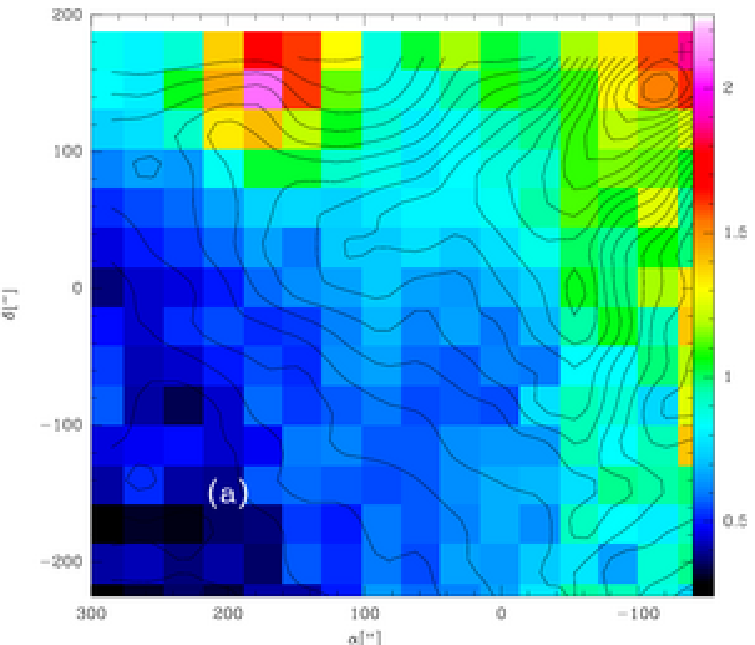}  
\includegraphics[angle=0,width=8.0cm,angle=0]{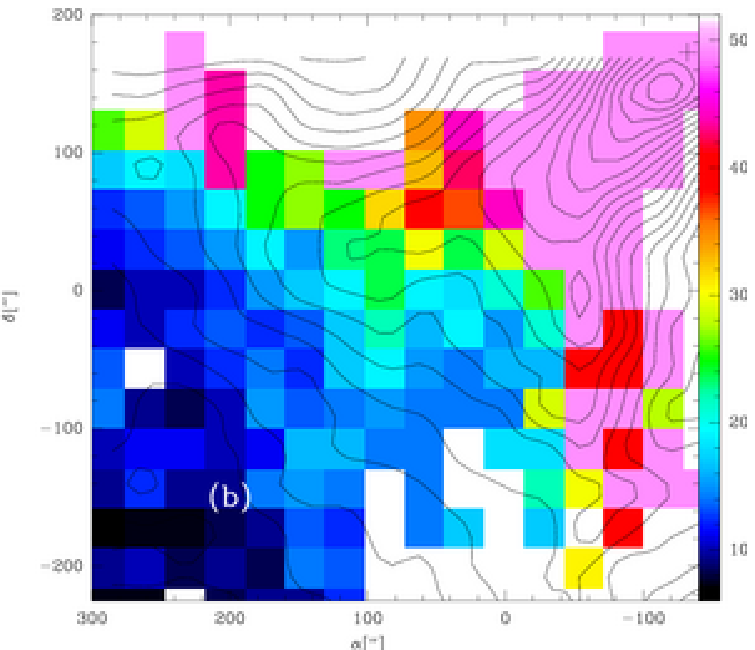}  
\caption {{\bf \em (a)}Distribution of observed \twCO4--3/\twCO1--0 ratio
in Cepheus~B.  {\bf \em (b)} Distribution of excitation temperatures
\ext\ derived from the $^{12}$CO 4--3/1--0 line ratio of integrated
intensities, assuming LTE. Contours in both panels correspond to
integrated intensities of \twCO\ 4--3.
\label{co43byco10}}  
\end{center}  
\end{figure*}   
 
Figure~\ref{co43byco10}(a) shows a map of the ratio of the
\twCO4--3/\twCO 1--0 intensities integrated over the velocity range
between -20 and -5~\kms\ for the region in Cepheus~B. It shows a smooth
gradient between ratios  of $0.3-0.6$ in the south-east and ratios of
$1.5-2.2$ to the north-west  of the mapped region.

Ratios of 0.9$\pm0.2$, as seen towards the bulk of the molecular
ridge, suggest that both transitions are thermalized and can be
explained in terms of LTE and optically thick emission. 
Neglecting the Rayleigh-Jeans correction, the ratio is expected to be
1.0 in this case.
The calibration uncertainty of line ratios is $\sim20$\%.  The
4--3/1--0 ratio is sensitive to excitation temperatures only
upto 50~K, beyond that the ratio remains almost constant with increasing
\ext.

While ratios less than 0.8 suggest sub-thermal excitation of \twCO 4--3,
ratios above 1.2 are inconsistent with above assumptions.  At the
position of the hotspot for which the spectra are shown in
Fig.~\ref{fig_posspec} 
the integrated intensity ratio  4--3/1--0 has a value of 1.9.  Similar
to this position, at all other positions with high 4--3/1--0 ratios, we
do not see any evidence of self absorption of either line by cold
foreground gas.  Within the framework of single \ext\  and LTE,
differences in optical thickness and volume filling factors of the
emitting regions of the two lines could still partially explain the
observed high 4--3/1--0 ratios.  However we consider strong outwardly
increasing temperature gradients to be the more plausible explanation.
Such a gradient results in an inconsistency in deriving a single \ext\
from this ratio, since the 4--3 transition will predominantly trace the
warm, still dense, outer regions, while the 1--0 traces the somewhat
inner, colder regions.  Such  high \twCO\ ratios are typically found in
the surface regions of molecular clouds subject to strong UV fields
\citep{castets1990,kramer1996,beuther2000,kramer2004}. 

Excitation temperatures derived from the 4--3/1--0 line ratios vary
smoothly between $\sim8$\,K in the south-east of the mapped region and
more than 50\,K in the north-west near the position of the hot core and
at the western edge of the molecular cloud (Fig.\,\ref{co43byco10}(b)).
In the following discussion, we have assumed a constant
excitation temperature of 50\,K  for 4--3/1--0$>0.9$.

\subsection{C$^0$/CO Abundances in Cepheus~B  
\label{sec_cico}}  
  
\begin{figure*}[htb]  
\begin{center}  
\includegraphics[angle=0,width=7.5cm,angle=0]{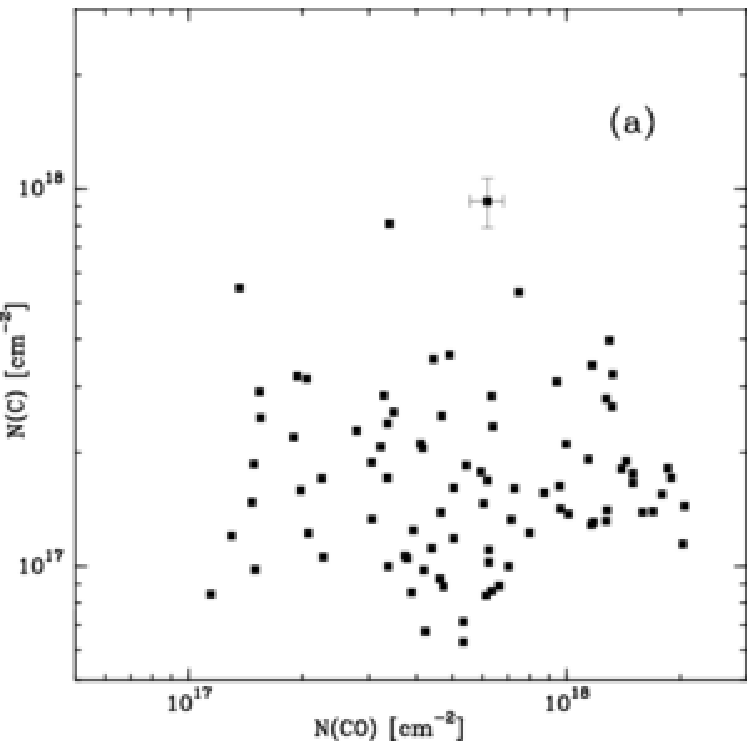}  
\includegraphics[angle=0,width=8.5cm,angle=0]{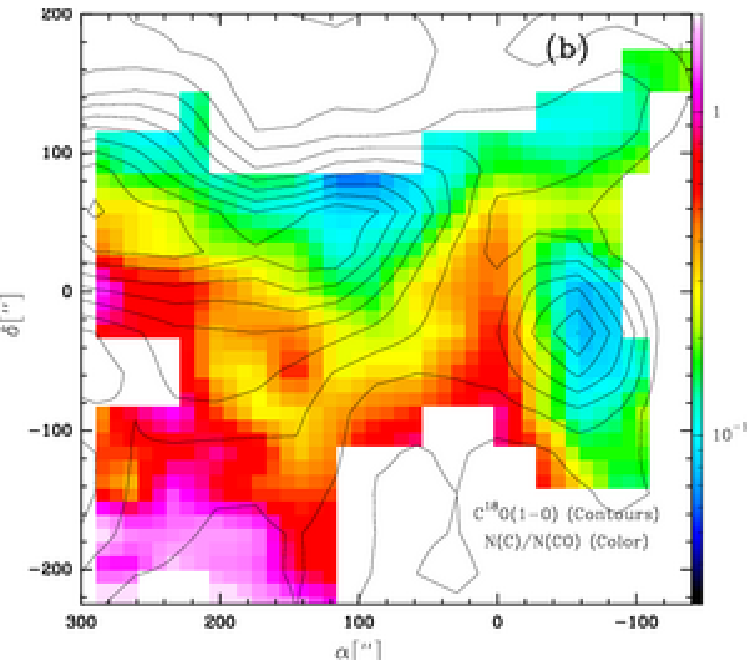}  
\caption {{\bf \em (a)} N(C) vs N(CO) for independent positions
(positions separated by 1\arcmin, i.e., beamwidth) in
the map. Typical
  errors of 15\% for N(C) and 10\% for N(CO) are shown for
  one position.  
{\bf \em (b)}The C$^0$/CO abundance ratio in Cepheus~B overlayed with the contours of  
integrated intensity of \CeiO1--0.
We have considered here only those positions for which we have \twCO 4--3 data
and for which the \CI\ and \CeiO\ intensities lie above 3 times their
respective rms noise levels.  
\label{cbycofig}}  
\end{center}  
\end{figure*}   
  
We have estimated the C$^0$ and CO column densities based on LTE
approximation. For these calculations we have used the observed \CI\ and
\CeiO\ 1--0 line intensities, the T$_{\rm ex}$ estimated from the
\twCO4--3/\twCO1--0 ratios and a relative abundance ratio [CO]/[\CeiO]
of 500 \citep{wilson1994}. Further we have assumed both \CI\ and \CeiO\
1--0 to be optically thin. The more detailed PDR analysis presented in
Sect.~\ref{sec_pdr}, strongly supports this latter assumption.

Assuming LTE N(CO) depends linearly on \ext\ above 20~K, while in
contrast, N(C) is rather independent of the assumed \ext\ to within
20\% between 20 and 150~K.  However, in between 10 and 20~K, N(C)
drops by more than a factor of 2, while N(CO) remains constant within
30\%.

Figure~\ref{cbycofig}(a) shows a scatterplot of the the C$^0$ column
densities, N(C) versus the CO column densities, N(CO). CO column
densities vary by more than one order of magnitude between $10^{17}$
and $2~10^{18}$~cm$^{-2}$, corresponding to optical extinctions
between $\sim1$\,mag and 30\,mag.  Although the range of values
of N(C) at different positions appear to be similar to the range
shown by N(CO), a histogram of N(C) shows that for most positions N(C)
does not show much variation, staying almost constant at a mean value of
$2.0\,10^{17}$\,cm$^{-2}$ with an rms of $1.4\,10^{17}$\,cm$^{-2}$.

Figure~\ref{cbycofig}(b) shows a map of the N(C)/N(CO) (\cbyco\ hereafter)
overlayed with contours of integrated intensities of \CeiO.
This plot shows that regions of higher \CeiO\
1--0 intensities, presumably tracing high H$_2$ column densities, show
rather low values of \cbyco, in contrast to the outskirts.  The ridge
and the south-western core show low values of \cbyco\ ($\sim 0.1$),
while the ratio rises gradually to values up to 4 outside these regions
where the CO emission is more diffuse.

\begin{figure}[h]  
\begin{center}  
\includegraphics[angle=0,width=8.0cm,angle=0]{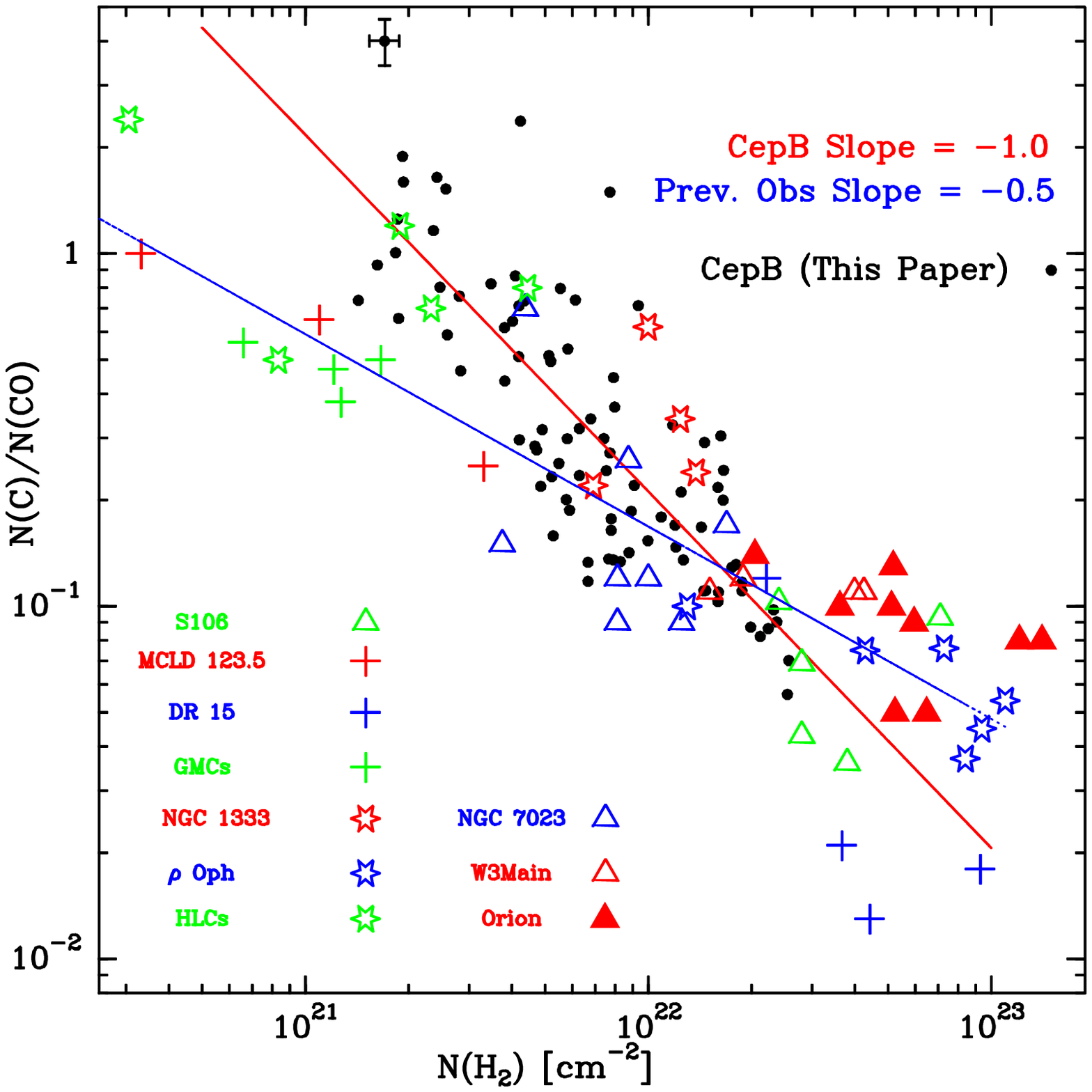}  
\caption {Column density ratio of C to CO in Cepheus~B.  
  Plotted as in Fig~\ref{cbycofig} are independent positions,
  which are  separated by 1\arcmin.  
  Also shown are observed values in other
  Galactic molecular clouds like NGC~7023 \citep{gerin1998}, S106
  \citep{schneider2003}, DR~15 \citep{oka2001}, MCLD 123.5+24.9
  \citep{bensch2003}, High Latitude Clouds \citep{ingalls1997}, NGC~1333
  \citep{oka2004}, $\rho$ Oph \citep{kamegai2003}, W3Main
  \citep{kramer2004} and Orion \citep{ikeda2002}. The points
  corresponding to GMCs are taken from \citet{plume1999} in which \CI\
  data of Cep~A, S~140, NGC~2024 and W3 were presented. The red drawn line
  corresponds to the parameter derived by fitting only the Cepheus~B
  data.  Typical
  errors of 15\% for N(C)/N(CO) and 10\% for \nhtwo\ are shown for
  one position. The blue line is the result of fitting data points
  for all previous observations.
\label{cepb_corr}}  
\end{center}  
\end{figure}

The constancy of C$^0$ column densities is also reflected in
Fig.~\ref{cepb_corr} showing a scatterplot of the C/CO abundance ratio
versus H$_2$ column densities or optical extinctions derived from
N(CO) assuming a CO/H$_2$ abundance of $8\,10^{-5}$ (Blake et al.
1987). The C/CO ratio is $\sim 1$ for 2~mag and drops to 0.1 for 20~mag.
A linear fit to the data results in

\begin{equation} 
  log\left[N({\rm C^0})/N({\rm CO})\right] = (-1.01\pm0.03) N({\rm
  H}_2) + (21.5\pm0.72)
\end{equation} 

with a correlation coefficient of -0.82.  The slope is almost $-1$ as
would be expected for nearly constant N(C).

Figure~\ref{cepb_corr} also shows the C/CO abundances derived in other
Galactic star forming regions and diffuse clouds. These studies cover a
range of optical extinctions between 0.3\,mag and $\sim100$\,mag with
\cbyco\ abundances varying by more than two orders of magnitude between
$\sim4$ and $\sim10^{-2}$, approximately following the slope we find in
Cepheus B.

Clearly, the \CI\ 1--0 line is not a straightforward tracer of optical
extinctions or total H$_2$ column densities and total masses.

The scatter of \cbyco\ abundance varies with A$_V$. For A$_V$'s upto
20~mag the scatter decreases from being almost comparable to the mean
\cbyco\ itself to about 50\% of the mean.  At $\sim20$\,mag the scatter
is the smallest, with the \cbyco\ abundance from five different data
sets agreeing to a value of 0.1 with a scatter of 0.05.  Note that
several studies which cover only a small range of extinctions find
rather constant C/CO abundances. 

In Sect.~\ref{sec_pdr} we show that clumpy PDR models naturally
explain the observed variation of the C/CO abundance.

\section{FUV intensity from FIR  continuum  
\label{sec_fuv}}  

The spectral lines observed in Cepheus~B which are used here are not
sensitive to the FUV radiation field, and there are no largescale \CII\
observations available in the literature for this region.  We thus
estimate the FUV radiation field from the observed FIR
continuum. Because dust grains absorb most of the FUV photons incident
on the cloud and most of that energy is reemitted in the form of
FIR continuum, and other heating mechanisms like cosmic ray
heating can be neglected, the intensity of the FUV radiation can be
inferred from the FIR continuum. This estimate is valid even if
a fraction of FUV photons escape without  
impinging the cloud. 
  
We have estimated the FIR intensity of the mapped region using the  
HIRES processed 60 and 100~\micron\ IRAS fluxes following the relationship  
\citep{nakagawa1998}:  
  
\begin{equation}  
I_{FIR} = 3.25 \times 10^{-11} \left[ \frac{f_\nu(60~\micron)}{\rm Jy~  
sr^{-1}}\right]  
+ 1.26 \times 10^{-11}\left [ \frac{f_\nu(100~\micron)}{\rm Jy~sr^{-1}}\right]  
\end{equation}

{The FUV intensity expressed in units of the Draine field
\citep{draine1978,draine1996} is given by  }
 
\begin{equation} 
 \chi= \frac{1}{1.71}\frac{4\pi}{(2\times1.6\,10^{-3})} I_{\rm FIR}
\end{equation} 
 
Following \citet{kaufman1999}, this formula takes into account  
that there is about equal heating of the grains by photons outside the FUV
band (leading to the factor 2) and 
that the bolometric dust continuum emission is a factor of $\sim2$ larger
than the FIR intensity derived above from the IRAS fluxes. 
We also assume that the emission stems from slabs which are illuminated
from all sides and the FIR emission stems from the back and front sides.
The latter two corrections approximately cancel out. 
Finally, in order to convert the FUV field expressed in units of Habing
field \citep[1.6$\times10^{-3}$ erg\,s$^{-1}$\,cm$^{-2}$\,sr$^{-1}$;][] {habing1968} to Draine field it is required to divide by
1.71.

\begin{figure}[h]  
\begin{center}  
\includegraphics[angle=0,width=8.0cm,angle=0]{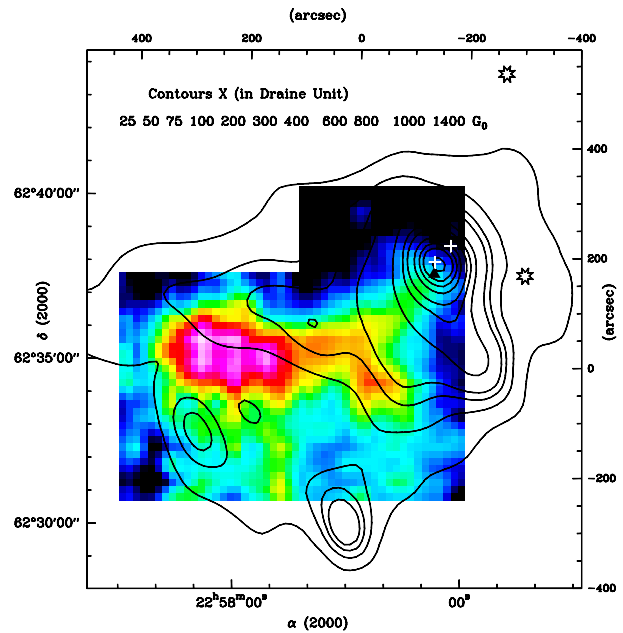}  
\caption  
{FUV intensity distribution (red contours) in Cepheus~B estimated from
the FIR  fluxes measured by IRAS, overlayed with \CI\ integrated
intensities (in  color). Contour  levels of the FUV intensity are
indicated at the top left corner. The  intensity scale for the \CI\
map are the same as in  Fig.~\ref{cepb_intmap}.
White stars denote HD217086 (north) and  HD217061, Crosses denote the
NIR sources A-NIR (south) and B-NIR and the  filled triangle denotes
the location of the {\em hot spot}.  
\label{cepb_fuv}}  
\end{center}  
\end{figure}   
  
Figure~\ref{cepb_fuv} shows the distribution of FUV intensity
calculated  as above, overlayed with the observed \CI\ emission.
Within the region mapped, the FUV field varies smoothly between
$\sim25$ in the south-east and 1500 times Draine field, in the
north-west near the hot spot. This variation of $\chi$ correlates well
with the corresponding variation of excitation temperatures derived
from the $^{12}$CO line ratios (Fig.~\ref{cbycofig}(b)). 
 
In Fig.~\ref{cepb_fuv} we have marked the positions of the nearest O and
B stars, HD217086 (north-west), HD217061 (west) and the embedded NIR
sources A and B (cf.  \citet{testi1995}).  
NIR-A, which does not have any infrared excess and which is deemed to be
a highly reddened main sequence star of spectral type B1-B0.5, is
responsible for the excitation of the associated blister \HII\ region
(the radio continuum source A).  The FUV intensity distribution
calculated as above is consistent with the results of \citet{testi1995}
and suggest that the star A-NIR is mostly responsible for the FIR
emission from the region.  The total estimated FIR luminosity from the
IRAS fluxes and for the assumed distance of 730~pc for Cepheus~B is
$\sim 10^4$~\lsun\ and this is consistent with the luminosity due to a
B1 type star as well.  
 
\section{PDR Model interpretation of C$^0$/CO vs N(H$_2$) relation  
\label{sec_pdr}}  

We interpret here the observed anti-correlation of the \cbyco\ abundance
ratio with the hydrogen column density (N(H$_2$)) in terms of PDRs
created due to the FUV irradiation of molecular material. For this we
have used the KOSMA-$\tau$ PDR model \citep{stoerzer1996,roellig2006}.
This model considers spherical clouds illuminated by an isotropic FUV
radiation field and cosmic rays, calculates the resulting chemical and
temperature structure of the cloud. This is followed by radiation
transfer calculations to estimate the emission of different species.
The KOSMA-$\tau$ model has \thCO\ chemistry included in its network. 
The emission from the models are calculated as a function of the
hydrogen volume density, FUV radiation field and mass of the spherical
clump. The KOSMA-$\tau$ model assumes that a clump has a power-law
density profile of $n(r)\sim r^{-1.5}$ for 0.2$\leq r/r_{cl}\leq 1$ and
$n(r)$ = const.  for $r/r_{cl}\leq 0.2$.  The surface density $n_{cl}$
is about half the average clump density.  
  
\subsection{Estimates of volume density  
\label{sec_dens}}  
  
\begin{figure}[h]  
\begin{center}  
\includegraphics[angle=0,width=9.0cm,angle=0]{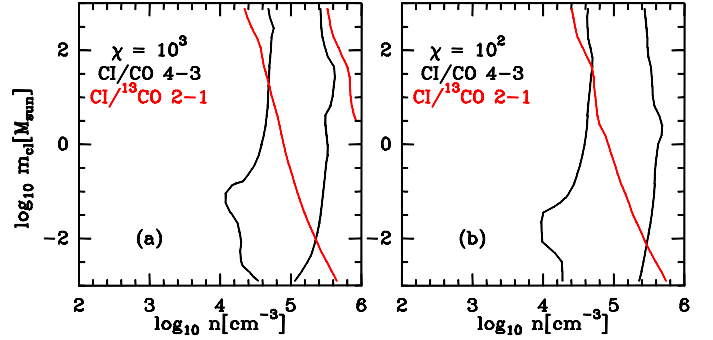}  
\caption{ 
  Results from the PDR models for external FUV fields of $\chi = 10^3$
  ({\bf a}) and $\chi = 10^2$, expressed in Draine units ({\bf b}). 
  Contours show integrated
  intensity line ratios of CI/\twCO 4--3 and CI/\thCO 2--1. The range of
  observed values is shown: 0.3--0.7 for  CI/\thCO\ 2--1, 0.2--0.6
  for CI/\twCO 4--3. 
  We consider only positions for
  which the C/CO abundance ratio has been estimated.  Both ratios
  suggest that the volume densities on an average lie between 10$^4$ and
  10$^5$~\cmcub. 
\label{fig_pdrdens} 
}  
\end{center}  
\end{figure}   
  
First, we have used PDR models to derive the range of allowed hydrogen
volume densities based on the observed CI/\thCO\ 2--1 and CI/\twCO
4--3 intensity ratios. At all the positions for which we have
calculated \cbyco\ abundance ratio the CI/\thCO\ 2--1 ratio varies
between 0.3 and 0.7, while the CI/\twCO 4--3 ratio varies between 0.2
and 0.6. Based on FIR maps the FUV field in the mapped region varies
between 50 and $10^3$  Draine units.  Figure~\ref{fig_pdrdens} shows the
contours bracketing the observed two line ratios for constant FUV
field and different PDR clump densities and masses.  These two line
ratios do not allow to discriminate between different clump masses.
The best fitting clump densities lie between $2\,10^4$ and
$8\,10^5$\,cm$^{-3}$, for clump masses between $10^{-2}$ and 10~\msun.
This range of densities is rather independent of the FUV field. 
We note that the clump masses are not very well constrained, however the
intersection of the two ratio contours in Fig.~\ref{fig_pdrdens}
provide physically reasonable limits on clump masses, which have been
loosely used in our subsequent analysis, but do not have any
significant impact on our final results.  The observed \twCO\ 4-3/1-0 line
ratios are consistent with this interpretation.

\subsection{Discussion  
\label{subsec_pdr}}

\subsubsection{Single Spherical PDR Model}

\begin{figure}[h]  
\begin{center}  
\includegraphics[angle=0,width=9.0cm,angle=0]{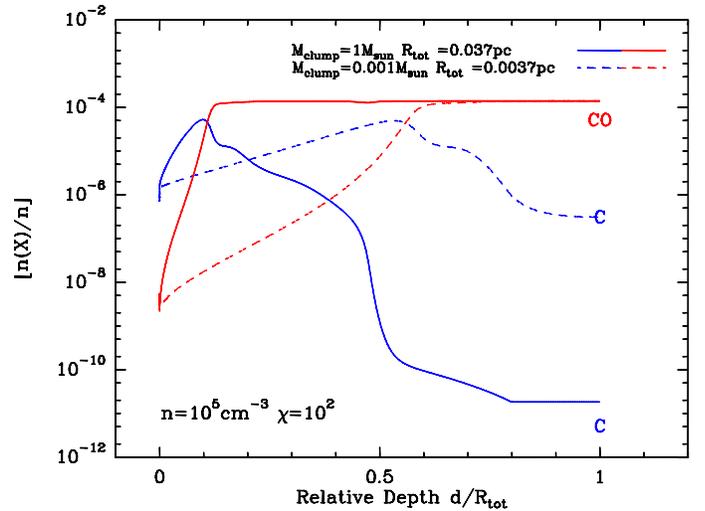}  
\caption{Abundance profiles of C$^0$ and CO calculated by the KOSMA-$\tau$
PDR model for two clumps with masses 1~\msun\ and 0.001~\msun.  The x-axis
is given in units of the relative depth $d/R_\mathrm{tot}$. A value of 1
denotes the center of the cloud. The FUV field strength is 10$^2$
in Draine units.
\label{abundprof} 
}  
\end{center}  
\end{figure}   

We first explore the single spherical PDR models described above, in
order to derive the observed \cbyco\ vs \nhtwo\ anti-correlation.  As shown
above, clumps with masses between 10$^{-2}$ and ~10 \msun\ with
densities $\sim 10^5$~\cmcub\ reproduce the range of \CI/\thCO 2--1
and \CI/\twCO 4--3 intensity ratios observed in Cepheus~B. In
Fig.~\ref{abundprof}, we present the calculated radial variation of
the C and CO abundances for two PDR clumps with masses of 10$^{-3}$ and 
1~\msun.  The surface density is fixed at $10^5$~\cmcub\ and the FUV
field is fixed at 10$^2$ (in Draine units), the average value
derived from the HIRES data in Sect.~\ref{sec_fuv}.

At the surface of the PDRs carbon is fully ionized by the incident FUV
radiation. Increasing depth leads to increased absorption of the FUV
radiation by both gas and dust. This allows C$^+$ to recombine with
electrons to form C$^0$. At even larger depths CO starts to form because
the photodissociating FUV photons are attenuated sufficiently leading to
a chemical stratification typical of PDR models, C$^+$/C$^0$/CO. The
absolute depth where this transition occurs depends on the gas density
$n$ and the FUV intensity $\chi$. In spherical models it also depends on
the size $R_\mathrm{tot}$ of the cloud since the local mean intensity
$J_\nu(r)=(4\,\pi)^{-1}\int\,I_\nu(r,\Omega)\,d\Omega$ at a certain
depth is larger for a small clump compared to a bigger one.  Accordingly
a clump with $M_{\rm cl}=1$~\msun\ consists of a much larger CO core
than a clump with $M_{\rm cl}=0.001$~\msun.  The average CO column
density increases with mass. The dependence of N(C) on the mass is
weaker. Once the clump becomes massive enough to significantly shield
the photodissociating FUV photons a further increase in mass does not
lead to larger C$^0$ columns but larger CO column densities.  This
behavior is illustrated in Fig.~\ref{abundprof} with the abundance of CO
at the inner core (with relative depth=1) of the clumps  being identical
in both cases.  Thus the 1~\msun\ clump has a much higher total fraction
of CO and lower \cbyco\ than the lower mass clump.  This indicates that
the lines of sight having high \cbyco\ abundance ratios are dominated by
clumps of smaller sizes.

\begin{table}[h]
\begin{center}
\caption{KOSMA-$\tau$ calculations of H$_2$, C and CO column 
densities of spherical clumps with n=10$^5$~\cmcub\ and $\chi =
10^2$ (Draine units). The column densities are for individual clumps, and
not the column densities observed per beam.
\label{pdrmod}}
\begin{tabular}{llllcc}
\hline\hline
M$_{\rm cl}$ & N(H$_2$)$_{\rm cl}$ & N(C)$_{\rm cl}$ & N(CO)$_{\rm cl}$ & $\frac{N(C^0)}{N(CO)}_{\rm cl}$ &
R$_{\rm cl}$\\
& 10$^{21}$ & 10$^{16}$ & 10$^{17}$ & &\\
\msun & cm$^{-2}$ & cm$^{-2}$& cm$^{-2}$& & pc\\
\hline
10    &  31  &  23  &  79  &  0.03  &  0.077\\
1     &  14  &  20  &  32  &  0.06  &  0.036   \\  
0.1   &  6.6 &  14  &  13 &   0.11  &  0.017    \\ 
0.01  &  3.3 &  8.7 &  4.0 &  0.22  &  0.008    \\ 
0.001 &  1.3 &  4.2 &  1.0 &  0.43  &  0.004    \\ 
\hline
\end{tabular}
\end{center}
\end{table}

Table~\ref{pdrmod} lists the clump-averaged C, CO and H$_2$ column
densities of the KOSMA-$\tau$ PDR model,  considering a single
spherical clump. We note here that the column densities in
Table~\ref{pdrmod} correspond to those calculated by the PDR model
for a single clump and do not corroborate with the observed column
densities per beam, as plotted in Fig.~\ref{cepb_corr}.  The \cbyco\ ratios
calculated by the PDR model is correlated with the mass of the clump.
We have checked the peak optical depths of the \CI\ line emission in a pencil
beam centered on the clump and for the line center position.
For all clumps studied here, these peak opacities are less than 0.5.
The clump and line averaged opacities of the \CI\ line are much less.
This justifies our previous assumption, used in the LTE analysis,
that the \CI\ line is optically thin. Similarly, the modelled $^{13}$CO peak
opacities indicate that the C$^{18}$O line is optically thin.

PDR models with n=10$^5$~\cmcub\  have clump sizes ranging between
0.004~pc (for m$_{\rm clump}$ = 10$^{-3}$~\msun) and 0.08~pc (for
m$_{\rm clump}$ = 10~\msun). In comparison, a beam of 60\arcsec\
corresponds to a diameter of 0.21~pc at the distance of Cepheus~B. It
is thus not possible to compare the observed \cbyco\ vs N(H$_2$) per
60\arcsec\ beam correlation directly with the N(H$_2$) per clump 
(N(H$_2$)$_{\rm cl}$ in Table~\ref{pdrmod}) obtained from the single
spherical PDR clumps.

\subsubsection{The Clumpy PDR Scenario}

The observed uniformity of the \CI\ intensity distribution throughout
the PDR (Fig.~\ref{cepb_intmap})  and the failure to interpret the
\cbyco\ vs \nhtwo\ using single spherical PDR model leaves open two
possible physical scenarios: (1) clumpiness of the medium, i.e.,
either there are multiple plane-parallel slabs stacked or there is an
ensemble of clumps along the line of sight, and (ii) a single,
homogeneous plane-parallel PDR slab inclined relative to the major
sources of UV radiation HD217086 and HD217061.

\begin{figure}[h]  
\begin{center}  
\includegraphics[angle=0,width=8.0cm,angle=0]{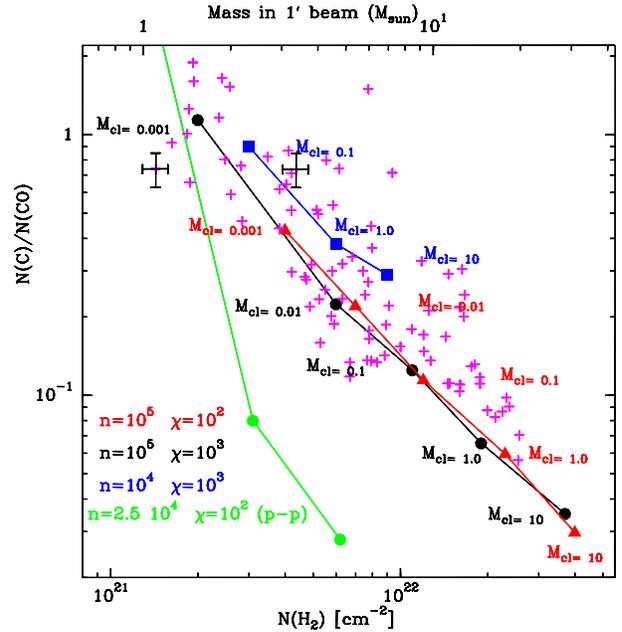}
\caption {KOSMA-$\tau$ PDR model interpretation of the observed
\cbyco\ vs N(H$_2$) anti-correlation. For comparison results of
face-on homogeneous plane-parallel models are also shown. Drawn lines
correspond to models the input parameters for which are marked at the
bottom left corner.  M$_{\rm cl}$ corresponds to mass of an individual
clump. M$_{\rm beam}$ shown along the upper x-axis is the mass of gas
in the 60\arcsec\ beam. The number of clumps in the beam is given by
M$_{\rm beam}$/M$_{\rm cl}$.
\label{correns}}  
\end{center}  
\end{figure}

We first consider the simplified clumpy PDR interpretation using
ensemble of clumps having the same mass and density at each position.
Figure~\ref{correns} shows that the observed \cbyco\ abundances and
total H$_2$ column densities of Cepheus B can be reproduced with
clumpy PDR models.  It shows results based on the ensemble
interpretation using the spherical PDR models with n=$10^5$~\cmcub,
$\chi =10^3$ and $10^2$ and n=$10^4$~\cmcub and $\chi =
10^3$. The drawn lines correspond to model calculations. We
consider an ensemble of clumps with mass M$_{\rm cl}$ (also marked in
Fig.~\ref{correns}) contributing to a total mass of M$_{\rm beam}$ in
a 60\arcsec\ beam. The total number of clumps in the beam is given by
N$_{\rm ens}$= M$_{\rm beam}$/M$_{\rm cl}$.  Table~\ref{tab_ens}
summarizes the defining parameters of the clump ensembles for which
the \cbyco\ abundance ratios are shown in Fig.~\ref{correns}.  We find
that the observed \cbyco\ anti-correlation depends primarily on the
characteristic mass of individual clump along a particular line of
sight and does not have a strong dependence on the assumed FUV field.
For clumps with densities of 10$^4$~\cmcub, masses below 0.1~\msun\
result in much higher values of \cbyco\ ratios than observed. 
The area filling factors ($\phi_{\rm A}$) of the ensemble vary between
1.6 and 3.3 (Table~\ref{tab_ens}), from the higher to the lower values
of clump masses considered in our calculations. This indicates that
the clumps fill the beam and in most cases partially overlap.

\begin{center}
\begin{table}[h*]
\caption{
\small Results of PDR model calculations for ensembles of clumps having
same mass and density. $\phi_{\rm A}$ is the area filling factor of the
clumps.
\label{tab_ens}}
\begin{tabular}{lllllll}
\hline
\hline
(n$_{\rm cl}$, $\chi$) & N(H$_2$)$_{\rm ens}$ & M$_{\rm ens}$ & M$_{\rm cl}$ & N$_{\rm ens}$ &
\cbyco$_{\rm ens}$ & $\Phi_A$\\
(\cmcub,$\ast$)  & cm$^{-2}$ & \msun\ & \msun\ &  &  & \\
\hline
                &  4.0~$10^{21}$  &     2.9 &    0.001 & 2859  &    0.43  &   3.3\\
                &  7.0~$10^{21}$  &     5.0 &    0.01 &  500  &    0.22  &   2.7\\
(10$^5$,10$^2$) &  1.2~$10^{22}$  &     8.6 &    0.1 &   86  &    0.11  &   2.2\\
                &  2.3~$10^{22}$  &    16.4 &    1.0 &   16  &    0.06  &   1.9\\
                &  4.0~$10^{22}$  &    28.6 &     10 &    3  &    0.03  &   1.6\\
\hline

                & 2.0~10$^{21}$  &     1.4 &    0.001 & 1429 &  1.14  &   1.7\\
                & 6.0~10$^{21}$  &     4.3 &    0.01 &  429 &  0.22  &   2.3\\
(10$^5$,10$^3$) & 1.1~10$^{22}$  &     7.9 &    0.1 &   79 &  0.13  &   2.0\\
                & 1.9~10$^{22}$  &    13.6 &    1.0 &   14 &  0.07  &   1.6\\
                & 3.7~10$^{22}$  &    26.4 &     10  &    3&   0.04  &   1.6\\
\hline

                &  3.0~10$^{21}$  &     2.1 &    0.1 &   21 & 0.9  & 2.4\\
(10$^4$,10$^3$) &  6.0~10$^{21}$  &     4.3 &    1.0 &    4 & 0.4  & 2.2\\
                &  9.0~10$^{21}$  &     6.4 &   10 &    1 & 0.3  & 2.5\\
\hline
\end{tabular}

$\ast$ In Draine units
\end{table}
\end{center}

 For the sake of completeness we also explored the second possible
physical scenario, an inclined plane-parallel PDR,  which might
produce the observed uniform PDR emission apparently deep into the cloud
and the observed \cbyco\ and \nhtwo\ anti-correlation.  For this
purpose we have used the plane-parallel PDR models illuminated from
one side by \citet{koester1994} in order to explicitly calculate the
\cbyco\ ratios and \nhtwo\ for selected values of volume densities and
UV-field. We have explored several values of $n$, slab-size (\nhtwo)
and $\chi$ and have found consistently that for face-on PDRs the
calculated \cbyco\ ratio drops much faster with \nhtwo\ as compared to
the observations. Figure~\ref{correns} shows the results of
calculation for a plane-parallel model with $n=2.5~10^4$~\cmcub\ and
$\chi=100$, for different slab-sizes or total \nhtwo. Changing the
inclination of the PDR has the same effect on both N(C) and N(CO), and
thus the \cbyco\ ratio remains unaltered, though \nhtwo\ changes as
the reciprocal of the sinus of the inclination angle. For this
particular set of models we find that at \nhtwo\ = 3~10$^{21}$~\cmsq\
the calculated \cbyco\ ratio is 0.08. In contrast, observationally we
find the \cbyco\ ratio to be 0.08 at \nhtwo\ $\sim 3~10^{22}$~\cmsq.
There is a discrepancy by a factor of 10 in \nhtwo, which can be
explained by an inclination angle of $6$\arcdeg. Similarly, different
inclination angles are required to bring the different plane-parallel
models to conform to the observed slope of the anti-correlation.
Hence, though it might be possible to explain individual positions by
making suitable choice of inclination angles, it is not possible to
reproduce the observed anti-correlation at all positions in Cepheus~B
using a single inclination angle.  Further, the anti-correlation as
has been shown in Fig.~\ref{cepb_corr} holds for a wide range of
sources so it is rather unlikely that all of them would either have
the same inclination angle or have different inclination angles and
still conform to the same anti-correlation.   The plane-parallel
models thus fail to reproduce the slope of the \cbyco\ vs \nhtwo\
anti-correlation entirely which extends over two orders of magnitude
in \nhtwo. Intuitively, both uniformity of \CI\ emission over the
entire ridge of Cepheus~B and the \cbyco\ vs \nhtwo\ anti-correlation
seen in most PDRs appear to be obvious outcomes of homogeneous
plane-parallel models which are inclined.  However, explicit
calculations suggest that although the homogeneous plane parallel
models predict almost constant N(C) with changing \nhtwo, they do not
reproduce the observed global \cbyco\ vs \nhtwo\ anti-correlation even
when inclined.  
 
We conclude that the spherical clumpy PDR model reproduces the
frequently observed global \cbyco\ vs \nhtwo\ anti-correlation over
the entire range of observed \nhtwo. These clumpy PDR models can thus
be used to understand the typical density and size (hence mass) of the
regions that dominate the observed emission. This does not imply the
absence of clumps of sizes smaller or larger than those reproducing
the observed \cbyco\ ratios.  However the emission cannot be
dominated by such clumps if the intensity and abundance ratios of
these clumps differ significantly from the observed numbers. Hence the
low observed relative abundance ratio \cbyco\ originate from positions
of high column density where the emission is dominated by a few more
massive clumps, while high ratios originate at positions covered by
many small clumps, having a low line of sight column density.

The calculations presented here are meant to be of qualitative nature in
order to derive insight into the clump parameters which determine the
observed \cbyco\ ratio. 

Figure~\ref{cbycofig} shows that the \cbyco\ ratio is higher
towards the south-eastern part of the ridge further away from the
sources of FUV radiation. This together with the results of clumpy PDR
models which find that higher \cbyco\ ratios are associated with
clumps of smaller mass present an interesting situation. In Cepheus~B
towards the north-western edge there are evidences of embedded sources
while there are none to the south-east.  Thus, using the clumpy PDR
models we see that regions with embedded sources correspond to lines
of sight with higher mass clumps.

\section{Summary}

We have presented the first largescale fully-sampled maps of \CI\
(492~GHz) and \twCO\ 4--3, of the edge-on PDR, Cepheus~B. The \CI\
emission does not delineate the PDR clearly, rather it traces the bulk
emission of the more quiescent part of the molecular ridge extending to
the east. Bulk of the PDR gas emits at velocities between -14 and
-11~\kms. The \twCO\ 4--3 emission is strongly dominated by the hotspot
present toward the north-western tip of the molecular cloud. We observed
\twCO 4--3/1--0 intensity ratios in excess of $1.5$ towards the
north-western edge of the cloud. Such ratios can only be explained in
terms of temperature gradients along the line of sight.  Excitation
temperature, derived from the observed \twCO\ 4--3/1--0 ratio, varies
between 20~K (in the quiescent parts of the cloud) to  50~K (close to
the hotspot). We conclude that \CI\ primarily traces the higher column
density regions while \twCO\ 4--3 traces the higher temperature and
density regions. 

In Cepheus~B we find that N(C) remains nearly constant although N(CO)
varies over one order of magnitude.  The observed \cbyco\ ratio varies
between 0.06 and 4.0 and shows a direct anti-correlation with the
hydrogen column density N(H$_2$). 

As estimated from the FIR  continuum emission, the FUV radiation field
in Cepheus~B varies between 25 \& 1500 in Draine units, with an
average value of 100 over most of the \CI\ emitting region. We
have used the KOSMA-$\tau$ PDR model to explain the observed intensity
ratios and the observed \cbyco\ vs N(H$_2$) anti-correlation.  These
PDR model calculations show that for $\chi = 100$, the observed
CI/\thCO2--1 and CI/\twCO4--3 ratios indicate local hydrogen densities
between 10$^4$--10$^5$~\cmcub\ with even higher values at a few
positions. The \twCO\ 4--3/1--0 ratio can be reproduced by the PDR
models and the modelled clumps indeed exhibit outwardly increasing
temperature gradients, as predicted from LTE analysis.

The observed global \cbyco\ vs \nhtwo\ anti-correlation and its slope
cannot be reproduced by single spherical PDR models or by
plane-parallel homogeneous slabs with or without inclination. We were
able to explain the observed anti-correlation between \cbyco\ and
N(H$_2$) considering an ensemble of clumps of the same mass and
density (n=10$^5$~\cmcub) for each observed position. We find that (i)
the \cbyco\ ratio is characteristic of the dominant mass in the
ensemble, clumps of smaller mass have higher \cbyco\ ratio and (ii)
clumps of larger mass have to necessarily dominate emission along
lines of sight with higher column density in order to reproduce the
observed lower \cbyco\ ratio.

\begin{acknowledgements}
We thank Hans Ungerechts for allowing the use of the datasets from the
Cepheus~B on-the-fly project at the IRAM 30m telescope.  This material is
based upon work supported by the Deutsche Forschungs Gemeinschaft (DFG) via
grant SFB494 and the National Science Foundation under Grant No. AST-0228974.
This research has made use of NASA's Astrophysics Data System.
\end{acknowledgements}

\end{document}